\documentclass[
aps,%
11pt,%
final,%
notitlepage,%
oneside,%
twocolumn,%
nobibnotes,%
nofootinbib,%
superscriptaddress,%
noshowpacs,%
centertags]%
{revtex4}
\usepackage[koi8-r]{inputenc}
\usepackage{multirow}
\begin{document}
\selectlanguage{english}
\title{Spectral and Photometric Monitoring of Distant
Core-Collapse Supernovae in the SAO RAS}

\author{\firstname{A.~S.}~\surname{Moskvitin}}
\affiliation{\saoname}

\author{\firstname{T.~A.}~\surname{Fatkhullin}}
\affiliation{\saoname}

\author{\firstname{V.~V.}~\surname{Sokolov}}
\affiliation{\saoname}

\author{\firstname{V.~N.}~\surname{Komarova}}
\affiliation{\saoname}

\author{\firstname{A.~J.}~\surname{Drake}}
\affiliation{California Institute of Technology, 1200 E.
California Blvd, CA 91225, USA}

\author{\firstname{R.}~\surname{Roy}}
\affiliation{Aryabhatta Research Institute of Observational
Sciences (ARIES), Manora Peak, Nainital, 263 129, India}

\author{\protect\linebreak \firstname{D.~Yu.}~\surname{Tsvetkov}}
\affiliation{Sternberg Astronomical Institute, University Ave. 13,
119992 Moscow, Russia}

\begin{abstract} This paper describes the aims, objectives and first results of the
observational program for the study of distant core-collapse
supernovae (SNe) with redshifts $z\lesssim0.3$. This work is done
within the framework of an international cooperation program on
the SNe monitoring at the 6-m BTA telescope of the Special
Astrophysical Observatory of the Russian Academy of Sciences, and
other telescopes. We study both the early phases of events (SN
type determination, redshift estimation, and a search for
manifestations of a wind envelope), and the nebular phase (the
effects of explosion asymmetry). The SNe, associated with cosmic
gamma-ray bursts are of particular interest. An interpretation of
our observational data along with the data obtained on other
telescopes is used to test the existing theoretical models of both
the SN explosion, and the surrounding circumstellar medium. In
2009 we observed 30 objects; the spectra were obtained for 12 of
them. We determined the types, phases after maximum, and redshifts
for five SNe \mbox{(SN 2009db, SN 2009dy, SN 2009dw, SN 2009ew, SN
2009ji).} Based on the obtained photometric data a discovery of
two more SNe was confirmed (SN 2009bx and SN 2009cb). A study of
two type II supernovae in the nebular phase (SN 2008gz and SN
2008in) is finalized, four more objects (SN 2008iy, SN 2009ay, SN
2009bw, SN 2009de) are currently monitored.
\end{abstract}

\maketitle

\section{INTRODUCTION}\label{Intro}
Core-collapse supernovae are usually considered as massive stars
(a star on the main sequence with more than 8 solar masses) at the
final stages of evolution, followed by a core collapse and a
powerful explosion (types Ib, Ic and II). An interest to these SNe
is driven primarily by the fact that they are suppliers of heavy
elements into the interstellar environment, and hence play an
important role in the processes of star formation in the galaxies.
In addition, compact relativistic objects are formed precisely in
the process of core-collapse SN explosions, giving a chance to
investigate the state of superdense matter, as well as the
conditions for the formation of such objects as pulsars and
microquasars.

Despite the ongoing active research of the SN phenomenon
worldwide, we are very far from understanding the processes before
and during the explosion. A new look at the problem appeared after
it was found that at least some Ib-c type SNe are associated with
cosmic gamma-ray bursts \mbox{(GRBs) \cite{Li08}}. This discovery
allows tracing the event straight from its onset, what previously
could only be done in isolated instances. For example, it was
discovered that at the onset of its explosion a SN can develop a
jet \cite{Fargion08}.

Nevertheless, there are many problems that still do not have any
decisive answers. We formulate some of these problems, and lay out
the possible solutions for the 6-meter BTA telescope of the
Special Astrophysical Observatory of the Russian Academy of
Sciences (SAO RAS) in Section \ref{Aims_set}. A specific
observational strategy of our program is considered in Section
\ref{Observations}. The main results of observations and
interpretation of the data obtained is presented in Section
\ref{Results}. The final Section \ref{Outro} discusses the plans
for future observations. Note that our research will mainly
concern the SNe with collapse of a massive core.

\section{PROBLEM DEFINITION}\label{Aims_set}

\subsection{Explosion Asymmetry}\label{Assymetry_expl}

Observations demonstrate that in a statistically significant
number of cases (30--40\%) an expansion of the SN envelope may be
asymmetrical (different from spherical) \cite{Kawabata02,
Trundle09,Maeda08,Modjaz08,Taubenberger09}.

Model calculations explain the observational data within the torus
or disk-shaped geometry of matter ejection during the explosion
\cite{Maeda02}. In this case the line profiles in the nebular
(more than 150 days after the maximum) SN stage serve as the
indicators. The following forbidden lines are usually used: the
[FeII] blend around 5200~\AA{}, [OI] 6300, 6363~\AA{} and
\mbox{[CaII]~ 7291, 7324~\AA.} An obvious problem needing to be
fully examined here is which mechanisms are responsible for the
explosion asymmetry. To answer this question, we perform spectral
observations at the BTA for a detailed study of line profiles. A
set of statistics of such events is very important to be able to
make any conclusions. An analysis of the existing data shows a
variety of possible explosion geometry variants
\cite{Taubenberger09}. Spectropolarimetric observations, now
possible with the BTA \cite{Afanasiev05} are as well challenging.
Such observations will expressly allow, firstly, to examine the
explosion geometry, and, secondly, to trace the distribution of
various elements in the outburst.

\subsection{Early Phase and Outburst Interaction with Surrounding Stellar
 and Circumstellar Matter}

In some rare cases, the observations of SNe can be performed at
the earliest phase: \mbox{SN 1993J \cite{vanDriel93},} SN 2006aj
\cite{Campana06}, SN 2008D \cite{Modjaz09}, SNLS-04D2dc and
SNLS-06D1jd \cite{Gezari08}, SNLS-04D2dc \cite{Schawinski08}. An
interpretation (and previously detailed modeling, see, \mbox{e.g.,
\cite{Blinnikov98}}) of the light curves in the ultraviolet,
visible light and X-rays allowed to discover an earlier predicted
effect of heating and acceleration of the progenitor star's
envelope by the shock wave and an egress of this shock onto the
surface of the star \mbox{(a ``shock breakout''
\cite{Colgate68}).} The most important effect of early
observations is a possibility of direct, outside the model
presentations, estimations of the size of the emitting region,
and, hence, the size of the presupernova \cite{Campana06}. It is
extremely important for the understanding of which stars go
supernova, as well as the mechanism of the explosion itself. It is
known that in the spectra of type I SNe, in contrast to type II,
there are no visible hydrogen lines. It is believed that the type
I SN progenitor stars during the evolution loose their hydrogen
envelopes, and in the case of  type Ic SNe, their helium envelopes
as well \cite{Nomoto_Mashimoto88}. It would be natural to expect
the manifestations of this envelope in the spectra, especially in
the early ones, as evidenced by Elmhamdi et al. \cite{Elmhamdi06}.
The authors interpret the absorption feature near 6300~\AA{} as a
shifted hydrogen line H$\alpha$ of the envelope, while their
method allows estimating the hydrogen mass as well.

Interestingly, the fact of existence of supernovae changing their
type in the course of time (e.g., SN1987K and SN1993J,
\cite{Elmhamdi06}) can be fairly easily explained within the
framework of a simple model. According to the present-day
perceptions, the most widely recognized type Ib-c SN progenitors
are: a) relatively low-mass stars \mbox{($8-20 M_{\odot}$)} in
binary systems (that loose their envelopes as a result of mass
transfer) \cite{Podsiadlowski92}, and b) Wolf-Rayet stars (loosing
their envelopes due to the stellar wind). A direct observational
task is to find the answer to the question: which evolutional
scenario of the type Ib-c SN progenitor stars is preferable? An
interpretation of the whole set of data (an estimation of
dimensions plus evaluation of mass of the hydrogen envelope) would
evidently be able to give a definite answer. In light of the
foresaid, a task for the BTA is the earliest possible spectroscopy
followed by mandatory monitoring of such events. We consider that
one of the most important components of this task is the detection
of the H$\beta$ line, which would become a decisive argument in
favor of the model discussed.

\subsection{A Wide SNe Luminosity Spread}

The brightness of type II SNe at maximum (for example, the events
SN 2005ap, SN 2006gy and \mbox{SN 2008es}) can reach $M = -22^m$
\cite{Drake08fz_09,Miller09}. However, it was found that the
luminosity of SNe of this type can be \mbox{$M_B = -14^m$} and
even fainter. In this respect, questions arise: how many such weak
flares occur, which percentage of them do we miss, and what effect
will the registration of such events have on the overall rate of
SN explosions in the galaxies?

In this case, another question seems natural: what is the
mechanism of powerful explosions and does it differ from the
``classical'' scenario? Within the framework of the stated
program, we carry out spectral monitoring of such events at the
BTA accompanied by multi-color photometry on the 1-meter
\mbox{Zeiss-1000} telescope of the SAO RAS and on other telescopes
in the context of international cooperation. The aim is to build
detailed light curves in different filters to be further used for
the estimates of mass and total energy of the progenitor, as well
as for a comparison of the observed characteristics (velocity,
line width, etc.) in the spectra of SNe with different
luminosities.

\begin{figure}
\setcaptionmargin{5mm} 
\onelinecaptionsfalse \captionstyle{normal}
\includegraphics[bb=40 130 575 630, clip, width=0.47\textwidth]{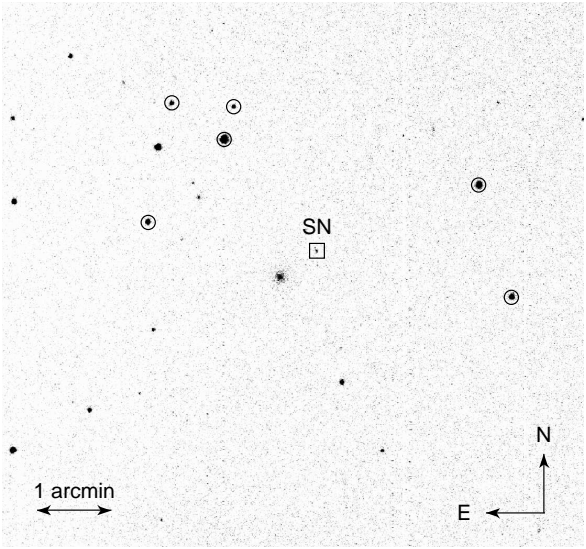}
\caption{Direct image of the CSS090216:100910+075434 (SN 2009bx)
field in the $V$ band, obtained with the \mbox{Zeiss-1000}
telescope on 27.840 February 2009 UT. In this and subsequent
figures the SN is marked with a square, while the circles encircle
the standards. At the time of observation the SN brightness
amounted to $B = 18.83 \pm 0.23$, $V = 18.90 \pm 0.11$, \mbox{$R_C
= 18.45 \pm 0.23$ \cite{CBET1744}.}
}\label{2009bx_V_chard}
\end{figure}

\begin{figure}
\setcaptionmargin{5mm} \onelinecaptionsfalse
    \includegraphics[bb=40 130 575 630, clip, width=0.47\textwidth]{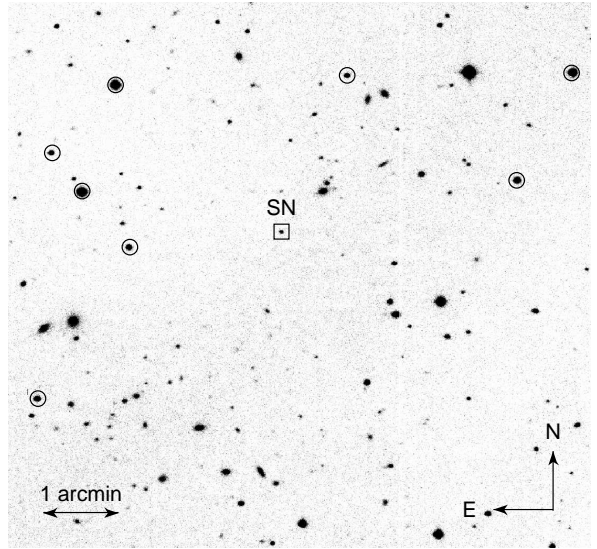}
   \caption{Direct image of the  CSS090319:125916+271641 (SN 2009cb) field in the $R$ band,
obtained with the Zeiss-1000 telescope on 29.987 March 2009 UT. At
the time of observation the SN brightness amounted to \mbox{$R_C =
19.78 \pm 0.10$ \cite{CBET1752}.}}\label{2009cb_R_chard}
\end{figure}

\subsection{Supernovae at Moderate and High Redshifts}

Important in understanding of both the phenomena of SNe and
gamma-ray bursts is a direct association of SNe with optical
afterglows of GRBs. However, spectroscopically such an association
was carried out only for the nearby events: \mbox{GRB 030329 /} SN
2003dh ($z=0.1687$), GRB 031203 / SN 2003lw ($z=0.1055$), and GRB
060218 / SN 2006aj \linebreak \mbox{($z=0.0335$)}~\cite{Kaneko07}.
Apparently, in the visible range such association can be reliably
done for the events at redshifts up to $z \sim 0.5$, when the
brightness of SNe (especially at peak brightness) will still
dominate in the overall radiation. It appears to be interesting to
compare the properties of SNe with and without GRBs. Here the
researchers are facing a problem: why are not all the nearby GRBs
associated with \mbox{SNe~ \cite{DellaValle06}?} Is this an effect
of observational selection linked with  the above mentioned
luminosity difference, or is it a distinction in the nature of the
phenomena? Within our observational program we perform
spectroscopic observations of distant SNe, where the monitoring of
events from their earliest phases is viewed essential. The most
important here are the measurements of velocities and widths of
the detected lines (the estimates of the total explosion energy),
and a comparison of the results obtained for the two events
studied (SNe associated with GRBs, and SNe not revealing this
connection).

It is obvious that the study of SNe at high redshifts is very
important for the understanding of the star formation history, in
particular, for an independent evaluation of its rate in the
Universe \cite{Dahlen04}, the evolution of the initial mass
function, etc. Modern specialized surveys can detect type IIn SNe
(the brightest in the ultraviolet) up to the redshifts
\mbox{$z\sim 2$ \cite{Cooke08,Cooke09}}. In this case a
perceivable task for the BTA would be multicolor photometric
observations, and construction of detailed light curves in the
context of international monitoring. From the data of broadband
photometry in four bands we may estimate the redshift and object
type (see, e.g., \mbox{\cite{Poznanski02,Kessler10}}).

According to the above, the main observational objective of the
program is spectral and photometric monitoring of SNe. Express
observations of SNe were recently made possible thanks to the
constantly updated and accessible on-line data, obtained in the
course of specialized sky surveys (see, e.g., \cite{Drake09}).

\begin{figure}
\setcaptionmargin{0mm} \onelinecaptionsfalse
\includegraphics[bb=24 19 707 532, clip, width=0.48\textwidth]{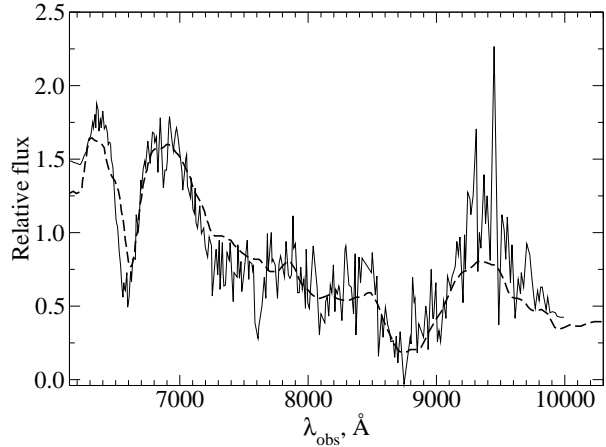}
\vspace{3.7mm}\caption{A comparison of the spectrum of
 CSS090319:142155+260102 (SN 2009db), obtained on April 3, 2009
with the BTA (the thin solid line) with the spectra from the SNID
database \cite{SNID}. The closest spectrum belongs to the type Ia
supernova SN 1999ee 11 days after maximum (the dashed line)
\cite{CBET1760}. An estimation of redshift from broad spectral
features: \mbox{$z=0.078\pm0.012$.}}\label{2009db_SNID}
\end{figure}

\begin{figure}
\setcaptionmargin{0mm} \onelinecaptionsfalse
    \includegraphics[bb=24 21 707 523, clip, width=0.48\textwidth]{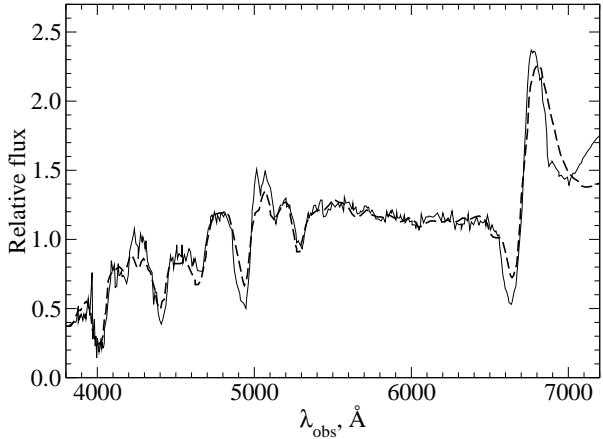}
\caption{A comparison of the spectrum
of CSS090421:133609+340319 (SN 2009dw), obtained on April 23, 2009
with the BTA (the thin solid line) with the spectra from the SNID
database. The closest spectrum belongs to the type \mbox{II-P}
supernova SN 2004et 15 days after maximum (the dashed line)
\cite{CBET1791}. Redshift estimation:  $z=0.042\pm0.003$. \mbox{SN
brightness} at the time of observation:
$R_C\approx19.0$.}\label{2009dw_SNID}
\end{figure}

\begin{figure}
\setcaptionmargin{0mm} \onelinecaptionsfalse
    \includegraphics[bb=24 21 707 532, clip, width=0.48\textwidth]{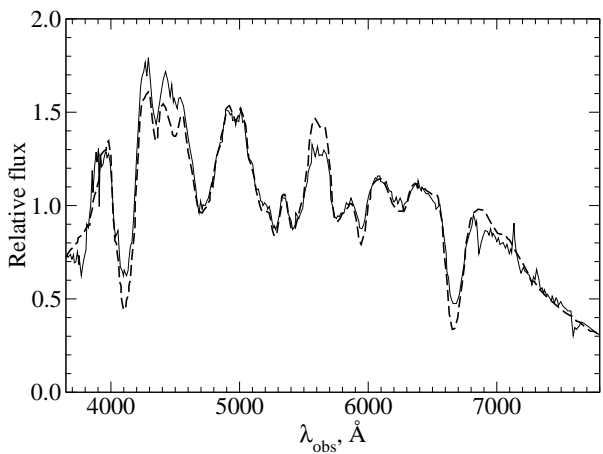}
   \caption{A comparison of the spectrum of CSS090422:150104+431314 (SN 2009dy),
obtained on April 24, 2009 with the BTA (the thin solid line) with
the spectra from the SNID database. The closest spectrum belongs
to the type Ia supernova SN 1994ae 6 days after maximum (the
dashed line) \cite{CBET1791}. Redshift estimation:
$z=0.089\pm0.003$. \mbox{SN brightness} at the time of
observation: $R_C = 18.63\pm0.20$, the USNO-B1 standards were used
for calibrations.}\label{2009dy_SNID}
\end{figure}

\begin{figure}
\setcaptionmargin{0mm} \onelinecaptionsfalse
    \includegraphics[bb=24 21 710 522, clip, width=0.48\textwidth]{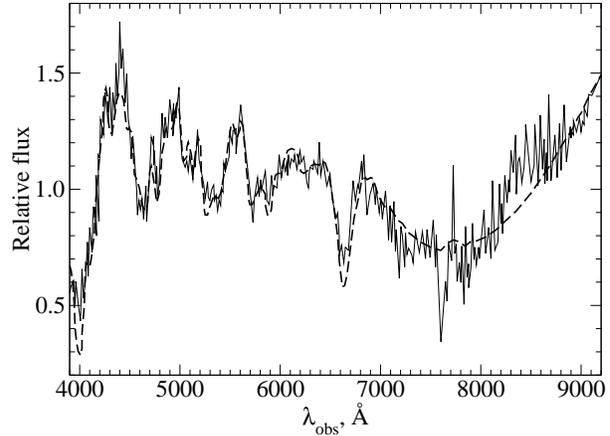}
   \caption{A comparison of the spectrum of CSS090516:163900+175858 (SN 2009ew),
obtained on May 17, 2009 with the BTA (the thin solid line) with
the spectra from the SNID database. The closest spectrum belongs
to the type Ia supernova SN 2003du 7 days before maximum (the
dashed line) \cite{CBET1815}. Redshift estimation:
$z=0.085\pm0.006$.}\label{2009ew_SNID}
\end{figure}

\begin{figure}
\setcaptionmargin{0mm} \onelinecaptionsfalse
    \includegraphics[bb=24 21 707 523, clip, width=0.48\textwidth]{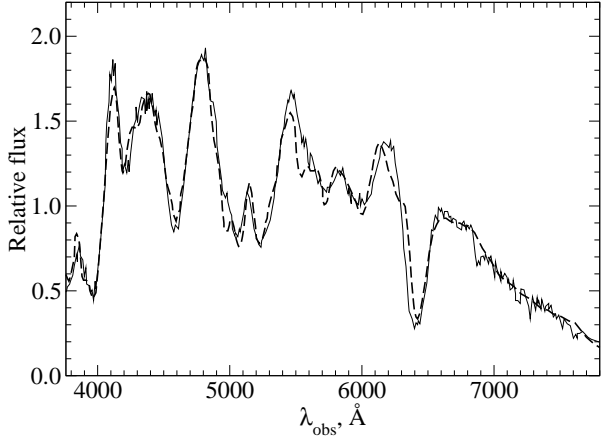}
   \caption{A comparison of the spectrum of CSS090923:155452+320506 (SN 2009ji),
obtained on September 25, 2009 with the BTA (the thin solid line)
with the spectra from the SNID database. The closest spectrum
belongs to the type Ia supernova SN 2003du 11 days after maximum
(the dashed line) \cite{CBET1960}.
  Redshift estimation: $z=0.048\pm0.004$.}\label{2009ji_SNID}
\end{figure}

\begin{figure*}
\setcaptionmargin{5mm} \onelinecaptionsfalse
    \includegraphics[bb=19 42 783 530, clip, width=0.9\textwidth]{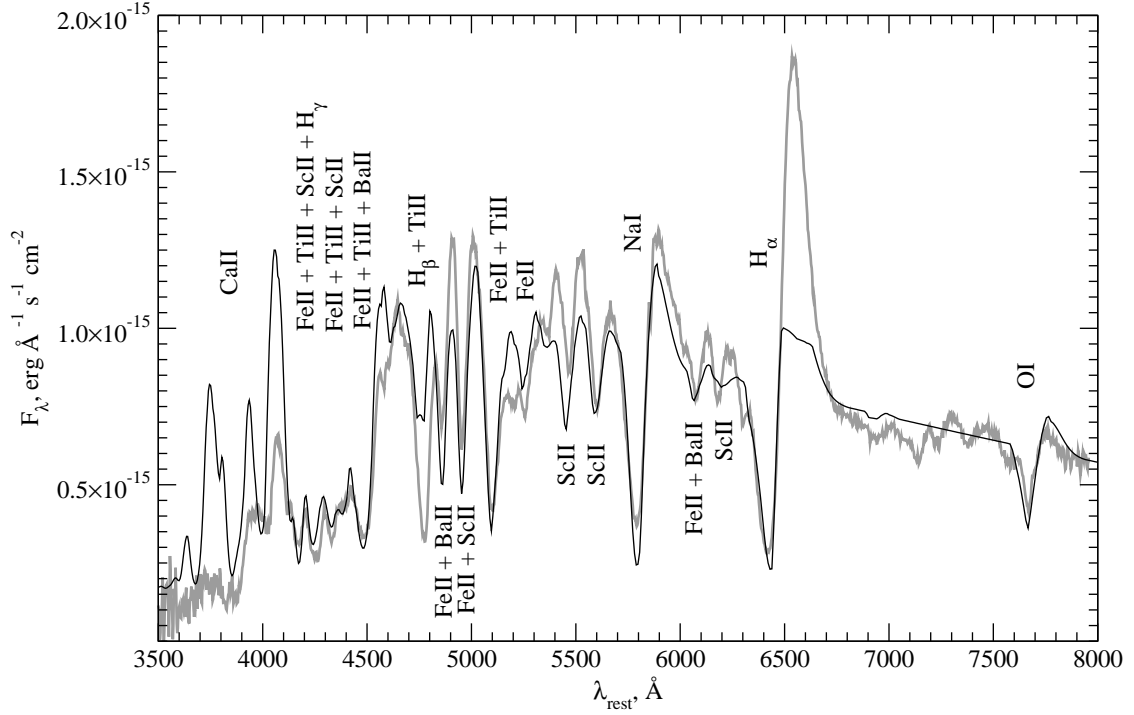}
   \caption{The spectrum of SN 2008gz, obtained on November 11, 2008 with the
   TNG+DOLORES telescope (the thick grey line). The thin line demonstrates the
   best fit by the model spectrum generated with the \texttt{SYNOW} code.}\label{08gz}
\end{figure*}

\begin{figure*}[tbp]
\setcaptionmargin{5mm} \onelinecaptionsfalse
    \includegraphics[bb=24 42 780 522, clip, width=0.9\textwidth]{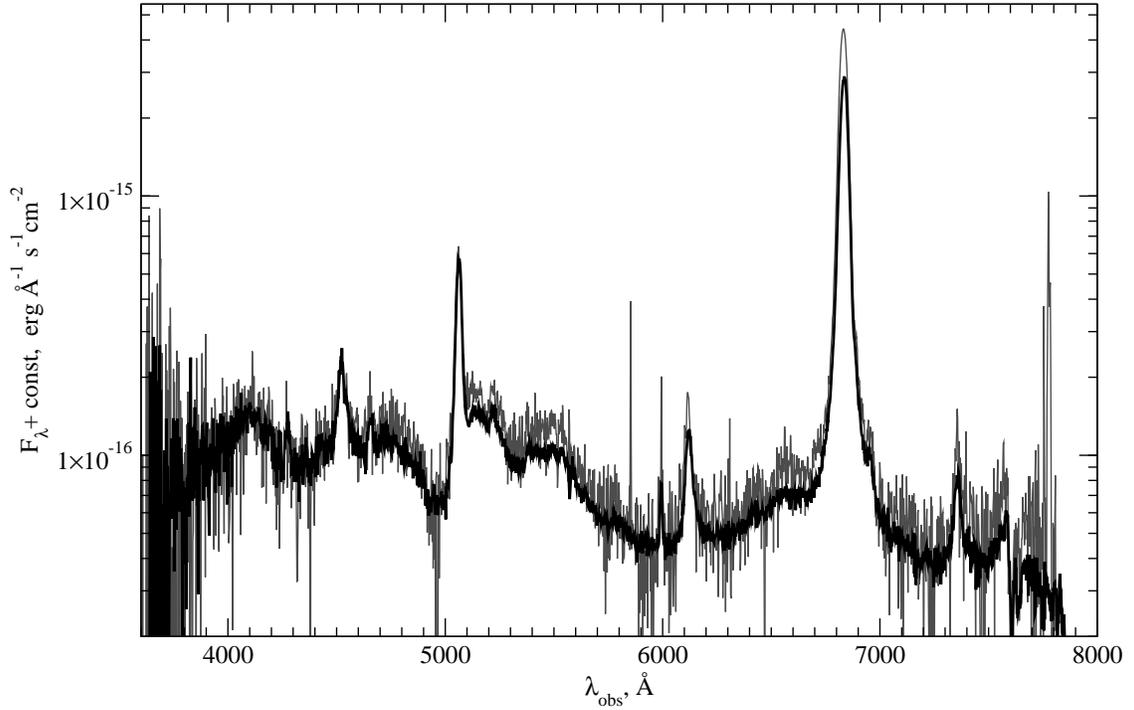}
   \caption{The spectra of SN 2008iy, obtained with the BTA+Scorpio on April 23
    (the black line) and September 25 (the grey line), 2009. The object's redshift,
    measured from the BTA spectra $z = 0.041$ is consistent with the data cited
    \mbox{in \cite{Miller09}}.
   }\label{08iy_sp}
\end{figure*}

\begin{figure*}
\setcaptionmargin{5mm} \onelinecaptionsfalse
    \includegraphics[bb=22 23 953 428, clip, width=0.9\textwidth]{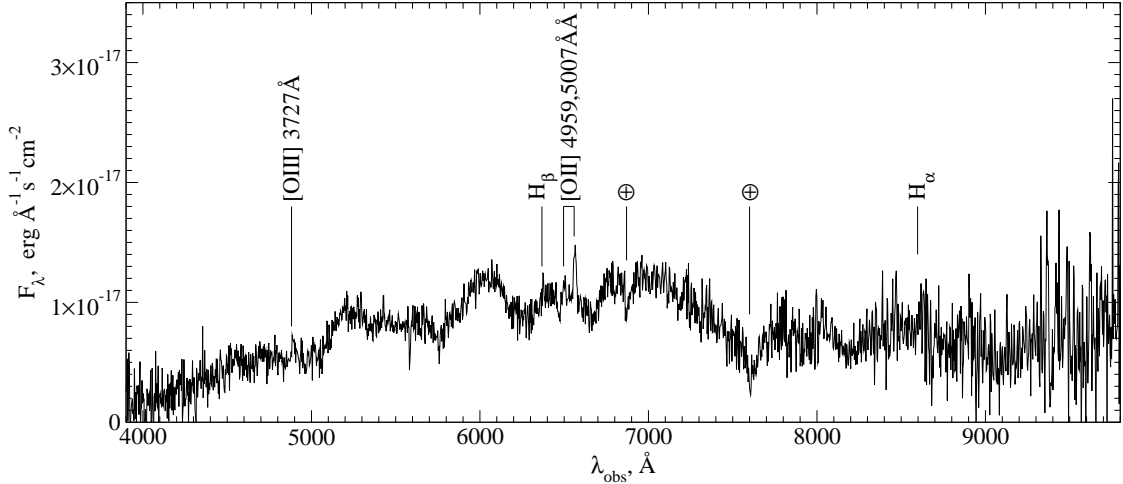}
\caption{Identification of the host galaxy emission lines in the
spectrum of SN 2009de, obtained with the BTA. We identified the
[OII] 3727~\AA, H$\beta$, [OIII] 4959, 5007~\AA~ lines, the traces
of the H$\alpha$ line were confirmed in the spectrum of the host
galaxy recently obtained with the Keck I telescope
\cite{Drake_private}. Redshift estimation, $z=0.311$, is close to
the value determined by fitting the broad features of the SN
spectrum with the SNID code.}\label{09de_lines}
\end{figure*}

\section{OBSERVATIONAL STRATEGY}\label{Observations}

An estimation of the number of the expected and accessible for
observations in the SAO events per year \mbox{($\delta
> -10^{\circ}$)} is: SN Ia---82, SN Ib---5, \mbox{SN Ic---8,} SN Ib-c---3,
SN IIn---11, SN IIP---10, other type II SNe---39 (to be precise
the evaluation was done for 2008 from the CRTS \cite{CRTS} data).
The classification of objects as core-collapse SNe is possible as
early as at the detection stage from the color \mbox{estimates
\cite{Drake09,Poznanski02,Kessler10}.} Since the main purpose of
the program is a study of precisely this type of SNe, we expect a
small percentage of type Ia SNe in our data. According to the
estimates of the detection rate we can expect about 1--2 events a
week.

The program observations are performed since 2009
in the framework of international cooperation for the studies of
core-collapse SNe on more than 10 telescopes in the USA, Italy,
Spain, India, Turkey and Russia.

In the SAO RAS the main observations are conducted on the 6-m BTA
telescope with a multimode device SCORPIO
\cite{Afanasiev05,Scorpio}, which is optimal for the program owing
to the capability of fast switching between the observational
modes. A large telescope is needed to achieve the required quality
of the data, as the expected apparent magnitudes of the objects
vary from $15^m$ to $25^m$. The observational strategy of the
program looks as follows:

\begin{enumerate}
 \item[1)] the early phase (the ``shock breakout''),
 \linebreak \mbox{$R = 15^m-19^m$:} spectroscopy with
  a resolution of $5-10$~\AA{} (VPHG550G, VPHG550R, VPHG1200R,
  VPHG1200G, VPHG1200B grisms \cite{Scorpio}), $UBVR_CI_C$ photometry;

 \item[2)] the early phase (SNe at moderate and high redshifts),
$R = 21^m-25^m$: $BVR_CI_C$ photometry, for the brightest
events---spectroscopy with a resolution of  $10-15$~\AA{}
(VPHG400, VPHG550G, VPHG550R);

 \item[3)] the late phase (nebular), \mbox{$R = 17^m-22^m$:} for bright events
($R = 17^m-19^m$) spectroscopy with a resolution of $5-10$~\AA{}
(VPHG550G, VPHG550R, VPHG1200R), $UBVR_CI_C$ photometry.
\end{enumerate}

The list of potential objects includes both the scheduled for
study and recently discovered events (the main sources of data are
the \mbox{CRTS \cite{CRTS},} \linebreak  \mbox{CBET \cite{CBET},}
and ATEL \cite{ATEL} catalogs). The selection of objects in a
particular night is determined by the task and conditions of
observations.

Data reduction is done in a standard way. The spectrum obtained is
compared with the SNID \cite{SNID} spectral library of the nearby
SNe and if its belonging to the SNe class is confirmed, then the
object type, its phase relative to the peak brightness, and
redshift are determined and published in the form of  \mbox{CBET
\cite{CBET}} or ATEL \cite{ATEL} telegrams. The spectra may be
used for a more detailed analysis, e.g. with the \texttt{SYNOW}
\mbox{code \cite{Branch2002}.} The photometric data can be used
for the absolute calibration of the spectra by flux, the
construction of light curves and for the estimations of physical
parameters of SNe.

\section{RESULTS}\label{Results}

The observational program is carried out since the first half of
2009. We perform the photometry of core-collapse SNe on the
Zeiss-1000 telescope of the SAO RAS as a follow-up program. In
2009, 30 objects were observed, for 12 of them the spectra were
obtained, and for 5 newly detected SNe (SN 2009db, SN 2009dy, SN
2009dw, SN 2009ew, SN 2009ji) we determined the types, the phases
after maximum, and redshifts. The discovery of two more supernovae
\mbox{(SN 2009bx, SN 2009cb)} was confirmed photometrically. We
completed the study of the nebular phase of two type II SNe (SN
2008gz and SN 2008in), the observations of four more objects (SN
2008iy, \mbox{SN 2009ay,} SN 2009bw, SN 2009de)~ are ongoing.

\subsection{Express Observations: Determination of Supernova Types and Redshifts}

The CRTS catalog \cite{CRTS} was used as a source of objects for
express observations. The belonging of two objects (SN 2009bx, SN
2009cb) to the SNe class  was tested photometrically at the
\mbox{Zeiss-1000} telescope (see Figs. \ref{2009bx_V_chard} and
\ref{2009cb_R_chard}). A transition from stellar magnitudes of the
standards in the $ugri$ system of the SDSS-DR7 catalog to the
$BVR_C$ system was made via the formulae from \cite{ugriz_UBVRI}.
For \mbox{SN 2009db,} \mbox{SN 2009dy,} SN 2009dw, SN 2009ew and
SN 2009ji we obtained the spectra, compared them with the SNID
database spectra, identified the SN types, estimated the phases
relative to the maximum, and redshifts from the broad spectral
features \mbox{(see Figs. \ref{2009db_SNID} --
\ref{2009ji_SNID}).}

\subsection{SN 2008gz and SN 2008in---Nearby Type II Supernovae}

In addition to the express observations of the newly discovered
SNe, the monitoring of the scheduled objects under study is
underway. In collaboration with the Indian and Italian members of
our international team  we traced the spectral evolution of SN
2008gz and SN 2008in. Light curves in the B, V, R, I bands  were
obtained. The bolometric light curve of \mbox{SN 2008gz} was
compared with the light curves of the same type SNe: SN 2004et and
\mbox{SN 1987A.} The explosion energy of \mbox{SN 2008gz} appeared
to be comparable with that of SN 2004et.  In the spectra close to
the maximum, the lines had the P\,Cyg profiles. They were studied
using the multiparameter \texttt{SYNOW} \mbox{code
\cite{Branch2002}.} The fitting result of the earliest spectrum of
SN 2008gz is presented in Fig. \ref{08gz} as an example. The
modeling revealed that the code restrictions are strong  for the
later spectra: the emission part of hydrogen profiles is poorly
described. To construct the curves of the envelope and photosphere
velocity drop for \mbox{SN 2008gz,} we measured the positions of
the absorption minima. All the details of the study of \mbox{SN
2008gz} are presented in \cite{Roy}.

\subsection{SN 2008iy---Type IIn SN or Quasar?}

One of the most interesting SNe studied in the framework of the
program is SN 2008iy
\cite{Drake08iy_08,Mahabal08iy_09,Catelan09_08iy}. It is
intriguing owing to the fact that its spectra, obtained at
intervals of about 5 months, changed very little (see Fig.
\ref{08iy_sp}). There comes a question on the nature of this
object. The light curve of SN 2008iy shows a record long
ascend---about a year in duration, and a very slow decline after
the maximum \cite{Miller09b}. The active phase of the outburst
lasts as long as several years. A decisive clue to the nature of
this object may be the detection of spectral features typical of
the nebular phase of SNe.

\subsection{Type II Supernovae SN 2009ay and SN 2009bw}

Along with our Italian and Moscow  colleagues we studied the
spectral evolution of type II supernovae SN 2009ay and SN 2009bw,
and noted an unusual behavior of brightness of these two type II-P
SNe at late phases (around the ``plateau'' on the light curves).
Extensive photometric data was obtained for both objects; for SN
2008bw we as well obtained the spectra at the BTA and other
telescopes.

\subsection{Cosmological Peculiar Type Ic Supernova \mbox{SN 2009de}}

SN 2009de was discovered in the context of the CRTS survey and was
studied photometrically and spectroscopically (Fig.
\ref{09de_lines}) on the telescopes of this survey, as well as
with the Palomar 60, Palomar 200, BTA, Zeiss-1000 and Keck I. The
study of such objects is important for the cosmological problems
of the program. In the near future we are planning to obtain deep
images on the BTA with the aim of detecting and studying the
spectral energy distribution of the \mbox{SN 2009de} host galaxy.
At the moment from the results of observations on the Zeiss-1000
and BTA telescopes it is known that the galaxy is weaker than $R_C
= 23.5$ and \mbox{$I_C = 24.0$.} However, the noisy spectrum
obtained with the Keck I showed the presence of the H$\beta$,
[OIII] 4958, 5006~\AA, and H$\alpha$ emission \mbox{lines
\cite{Drake_private},} suggesting the possibility of obtaining
with the BTA of the photometric data for modeling the spectral
energy distribution of the host galaxy.

\section{CONCLUSION}\label{Outro}

We suppose that the data obtained in the context of our
international monitoring will significantly improve the
understanding of a yet largely mysterious connection of gamma-ray
bursts with core-collapse supernovae
\cite{Sonbas2008,Moskvitin2010}. In an attempt to answer the
questions on the nature of the progenitors and the explosion
mechanisms of SNe and GRBs, as well as about their similarities
and distinctions, it is necessary to conduct observations of both
the most early phases, closest to the onset of the explosion, and
the later phases (a link with asymmetry). Statistical estimates of
the rate of SN and GRB explosions, and a comparison of these data
with the star formation rates play an important role. Hence a
separate objective emerges to study the regions of host galaxies
and the host galaxies themselves that are undergoing these
explosions.

Given the characteristic times during which an almost complete
fade-out of the SN brightness occur, we can say that the program
for studies of distant core-collapse SNe at the BTA has just
begun. At this stage, in addition to the early observations of new
objects, we have to continue the observations of individual
objects for their detailed study. In particular, it is important
to study the properties of the host galaxies of SN 2008iy and SN
2009de with the means of broadband photometry.

\begin{acknowledgments}
The authors are grateful to the program co-applicants and
colleagues involved in the international collaboration:
T.\,N.~Sokolova, G.\,M.~Beskin, V.\,L.~Plokhotnichenko,
S.\,V.~Karpov, V.\,P.~Goranskij, O.\,I.~Spiridonova,
A.\,F.~Valeev, O.\,N.~Sholuhova, \linebreak S.\,N.~Fabrika,
A.\,N.~Burenkov (SAO RAS), I.\,M.~Volkov (SAI); E.\,Sonba\c{s}
(University of Adiyaman, Tur- key); A.\,J.~Castro-Tirado,
J.~Gorosabel (IAA---CSIC, Spain); S.\,B.~Pandey, Brajesh Kumar
(ARIES, India); C.~Inserra, S.~Benetti (INAF, Italy), as well as
the reviewer for the comments which helped improve the
presentation of material. This work was supported by the grant RNP
2.1.1.3483 of the Federal Agency of Education of Russia and by the
grant of the Russian Foundation for Basic Research (project no.
\mbox{10-02-00249a}).
\end{acknowledgments}

\end{document}